# Passive Temperature-Compensating Technique for Microstructured Fiber Bragg Gratings

Minh Châu Phan Huy, Guillaume Laffont, Véronique Dewynter, Pierre Ferdinand,
Dominique Pagnoux, Bernard Dussardier, and Wilfried Blanc

*Abstract*— **The thermal drift of the characteristic wavelength of fiber Bragg gratings (FBGs) photowritten in the core of microstructured fibers (MOFs) is significantly reduced by inserting a liquid of suitable refractive index into their holes. For instance, the spectral range of variations is divided by a factor of 4 over a temperature range larger than 20°C in a 6-hole MOF, and the maximum sensitivity is reduced. Such passive FBG temperature compensation technique is of great interest for applications involving accurate sensing free of thermal effects.**

## I. Introduction

FIBER Bragg gratings (FBGs) consist in a permanent longitudinal modulation of the refractive index of the core of an optical fiber, that is commonly achieved by making the fiber photosensitive and then by exposing it to a UV interference pattern [1]. Due to the modulation, the forward propagating mode is coupled to a backward propagating mode at the resonant wavelength $\lambda_B$ (the so-called Bragg wavelength) given by the well known relation $\lambda_B = 2 \cdot n_{eff} \cdot \Lambda$ , where $\Lambda$ is the period of the longitudinal refractive index modulation and $n_{eff}$ is the average effective index of the forward mode. Over the twenty past years, FBGs have been largely used as narrow bandwidth reflectors in many applications, in particular in the fields of telecommunications, fiber lasers and sensor systems [2].

FBGs can be used as transducers for many sensing applications because both $\Lambda$ and/or $n_{eff}$ vary according to external conditions such as applied strain. Thus, one can take advantage of these variations to determine the change in the measurand in an indirect way, by detecting the related change in $\lambda_B$. One necessary condition for providing an accurate and reliable measurement is that $n_{eff}$ and $\Lambda$ do not change with other parameters, and in particular with temperature. Unfortunately, in standard fibers, both $\Lambda$ and $n_{eff}$ are also sensitive to temperature. The change of the period $\Lambda$ is due to the thermal expansion of the fiber whereas the effective index of the mode experiences the change in the refractive index of the silica arising from the thermo-optic effect [3]. In standard germano-silicate single-mode fibers, the cross-sensitivity of $\lambda_B$ with respect to temperature is typically 10 to 13 pm/°C at room temperature [4].

Recently, FBGs have also been photowritten in the core of microstructured optical fibers (MOFs) that guide light thanks to the modified total internal reflection principle [5]. These fibers are made of a strand of silica glass with a lattice of micrometric air channels running along their length and characterized by their diameter d and the distance between neighbouring centers, so-called "the pitch" D. The core is constituted by one (or several) missing hole(s) in the center of the holey lattice that acts as the optical cladding. The propagation characteristics of MOFs, which have been largely explored since their invention by Russell *et al.* in 1996 [6], have attracted a large interest for numerous applications, in particular in telecommunications, nonlinear optics and high optical power transmission [7]. Another attractive feature of this kind of fiber is that one can benefit from the interaction between the guided light and a foreign medium inserted into the holes of the cladding, for sensing applications [8]. In particular, the effective index of a guided mode can be significantly modified when the refractive index of this medium is changed, and FBGs photowritten in the core of a MOF may be used as transducers for high sensitivity refractive index measurements [9], [10]. However, as in standard fibers and for the same reason, the resonance wavelength of FBGs in empty MOFs depends also on the temperature. The temperature sensitivity $S = \Delta\lambda_B/\Delta T$ varies from 10 pm/°C to 20 pm/°C, depending on the core constituents [11]. This

Manuscript received June 30, 2007. This work has been co-funded by the French Ministry of Research (ACI Nouvelles Méthodologies Analytiques et Capteurs) and by the Institut National de Recherche et de Sécurité (INRS).

M.C. Phan Huy was with the Commissariat à l'Energie Atomique, Centre d'Etude de Saclay, 91191 Gif-sur-Yvette (France). She is now with the Laboratoire de Photonique et de Nanostructures, UPR CNRS 20, route de Nozay, 91460 Marcoussis, France (e-mail: minh-chau.phan-huy@lpn.cnrs.fr)

G. Laffont, V. Dewynter and P. Ferdinand are with the Commissariat à l'Energie Atomique, Centre d'Etude de Saclay, 91191 Gif-sur-Yvette, France (e-mails: guillaume.laffont@cea.fr, veronique.dewynter@cea.fr, and pierre.ferdinand@cea.fr)

D. Pagnoux is with the Photonics Department, Xlim Institute, UMR CNRS 6172 and Université de Limoges, 123 avenue Albert Thomas, 87060 Limoges cedex, France (corresponding author phone: +33-555-457-247 ; fax: +33-555-457-253 ; e-mail: dominique.pagnoux@xlim.fr).

B. Dussardier and W. Blanc are with the Laboratoire de Physique de la Matière Condensée, UMR CNRS 6622 and Université de Nice Sophia Antipolis, Parc Valrose, 06108 Nice, France (e-mails: bernard.dussardier@unice.fr and wilfried.blanc@unice.fr).

sensitivity can be a serious problem in applications in which the resonance wavelength of the FBG is expected to remain stable with temperature, as in wavelength-division multiplexing (WDM) devices for telecommunication systems, or in narrow-bandwidth fiber lasers or in sensors designed for accurate strain or pressure measurements.

Several techniques have already been proposed to passively stabilize the resonance wavelength of photowritten FBG *vs* temperature. One consists in compensating the temperature-induced variation of the resonant wavelength by varying the strain in the grating. This can be achieved by bonding the FGB under tension on a passive device comprising two materials with different thermal-expansion coefficients [12], [13]. One can also make use of a special glass material with a negative thermal expansion coefficient [14]. A reduction of the shift of the Bragg wavelength to less than 1/20 of that of an uncompensated FBG has thus been obtained within a 70°C temperature range [13].In an alternative technique, the uncoated strained fiber with the FBG is bonded with epoxy resin into a liquid crystalline polymer tube that has a negative thermal expansion coefficient [15]. By this mean, the shift of the Bragg wavelength can be divided by 8 over 100°C [15] These techniques are efficient but they are difficult to implement. Moreover, they require specific packaging and/or gratings mounted under tension, rendering them unsuitable for sensing applications. In this paper, we propose a simple passive method to efficiently stabilize $\lambda_B$ of MOFs designed for sensing purposes, over a significant temperature range. The principle of this method, based on the use of specific index matching liquids inserted into the holes, is first described. Then, experimental results obtained with two different MOFs are reported and discussed.

## II. PRINCIPLE OF THE METHOD AND EXPERIMENTAL IMPLEMENTATION

In FBGs photowritten in standard fibers, or in empty MOFs, the temperature dependent spectral shift of $\lambda_B$ is only due to thermal effects on silica, *i.e.* the thermal expansion of the material and the increase of its refractive index. Consequently, $\lambda_B$ experiences a red shift as temperature rises. In order to stabilize $\lambda_B$ *versus* temperature, the effect of the guided mode effective index should counterbalance that of the period $\Lambda$. In other words, $n_{eff}$ should decrease as the temperature rises. Obviously, this condition cannot be fulfilled in standard fibers because of the thermo-optic properties of doped or pure silica. However, in the case of MOFs, there is an opportunity to reach this goal as the propagating field also interacts with the medium inserted into the holes whose refractive index influences the effective index of the mode.

Let us specify a little more the thermal dependence required for the refractive index $n_l$ of the medium. To this end, we must remind that the equivalent index of the heterogeneous cladding $n_{clad}$ is the effective index of the fundamental mode able to exist in this cladding whose extension is supposed to be infinite. $n_{clad}$ is given by the following relation [16] :

$$n_{clad}^2 = \frac{\iint n^2 \cdot |E|^2 \, dS}{\iint |E|^2 \, dS} - \frac{\iint \left|\frac{dE}{dr}\right|^2 \, dS}{k^2 \cdot \iint |E|^2 \, dS} \quad (1)$$

where E is the electric field; n is the index of the silica or of the medium filling the holes, depending on the considered point of the cross-section of the cladding; k is the modulus of the wave vector in the vacuum; S is the area of an elementary cell of this cross-section; and r is the distance to the center of the fiber.

In its turn, the effective index of the guided mode $n_{eff}$ is approximately a weighted average of the cladding equivalent index and the silica core index $n_s$. The weighting coefficient is the fraction of the energy of the mode located in each region. From these qualitative considerations, one can easily deduce that $n_{clad}$, and consequently $n_{eff}$, are decreased as $n_l$ is decreased.

Finally, the required decrease of $n_{eff}$ as the temperature is increased can be obtained by filling the holes with a liquid whose refractive index is a decreasing function *vs* temperature. In this paper, we propose to stabilize the resonance wavelength of a FBG photowritten in MOFs by filling the holes with such a liquid.

Two different Ge-doped core MOFs (called Fiber A and Fiber B, respectively) have been manufactured at the Xlim Institute by means of the usual stack-and-draw technique [6]. The core of each perform was made from Ge-doped rod whose refractive index $n_{core}$ is ~ 8 $10^{-3}$ higher than that of pure silica. Each Ge-doped rod has been extracted from a MCVD preform manufactured at the University of Nice (LPMC laboratory) by means of mechanical machining followed by HF acid attack. The preform of Fiber A was a 18-hole structure with a two-ring triangular pattern of respectively 6 and 12 silica capillaries, surrounding the Ge-doped rod, stacked into a maintaining silica tube. The perform of Fiber B was a 6-hole structure with only one ring of 6 silica capillaries around the core. Each fiber was then drawn by means of a conventional drawing tower. The cross sections of the fibers are shown in Fig. 1. Fiber A has a 5 μm photosensitive core surrounded by 18 holes with a diameter d ~ 3.1 μm and a pitch D ~ 6.7 μm (Fig. 1a) whereas Fiber B has a 11μm photosensitive core surrounded by 6 large holes with a diameter d ~ 15 μm and a pitch D ~ 16 μm (Fig. 1b).

The FBG photowritting was performed at CEA LIST laboratories in Saclay, using a Lloyd mirror interferometer setup including a CW frequency-doubled argon 'Fred' laser emitting at 244 nm. Prior to photowriting, the MOFs were $H_2$-loaded at 160 bar and 25°C for one week to enhance the photosensitivity of their Ge-doped core. The resonance wavelength of the fundamental mode was chosen to be close to 1550 nm at 25°C. A typical transmission spectrum of a FBG photowritten in Fiber A is shown in Fig. 2. The full width at half maximum (FWHM) is approximately 0.25nm.

In order to measure the variations of $\lambda_B$ *versus* temperature in the MOFs, we used the experimental setup depicted in Fig. 3. The region of the MOF where the FBG was written was set in a cell which temperature is regulated by a Peltier element. The light from a tunable narrow bandwidth laser source (1 pm resolution) was launched into the empty MOF (Fig. 3, configuration ① ) through a 50/50 fiber coupler. The light reflected by the FBG in the MOF and by the two reference FBG (used to provide accurate wavelength references) was detected and registered *versus* wavelength at different temperatures from 10°C to 33°C. The experimental shift of $\lambda_B$ was ~ 10 pm/°C over this temperature range for both Fiber A and Fiber B.

In the next step, several index liquids (perfluorocarbon oils from Cargille Labs Inc.) with low viscosity were successively inserted by means of capillary force into the holes of the MOF (Fig. 3, configuration ② ). The refractive index of each oil and its thermal sensitivity is synthesized in Table I. At last, the filled fiber was again set in the temperature regulated cell (Fig. 3, configuration ③, same as configuration ① ) and the shift of $\lambda_B$ *versus* temperature was measured, following the same process as for the empty fiber. The experimental results obtained with the different index liquids are reported and discussed in the next section.

### III.  EXPERIMENTAL RESULTS AND DISCUSSION

In order to implement the above principle, the index liquids were chosen due to their negative refractive index sensitivity *versus* temperature around 25°C. An other condition was that their refractive index $n_l(T)$ had to remain lower than a ceiling value $n_{max}(T)$, in order to make sure that the equivalent index of the cladding, so-called $n_{clad}(T)$, verifies the relation $n_{clad}(T)<n_{core}(T)$ over the whole range of temperature. This condition must be fulfilled so that light remains guided into the core thanks to the modified total internal reflection principle. The refractive indices of the different liquids at 1550 nm and 25°C, denoted $n_{ref}$, were provided by the manufacturer. They take values from 1.3897 to 1.4458, depending on the liquid (Table I). We observed that their thermal sensitivities $dn_l/dT$, also provided by the manufacturer, remain very similar: between $-3.96 \cdot 10^{-4}$ and $-4.12 \cdot 10^{-4}$ refractive index unit (r.i.u.)/°C from 15°C to 35°C.

The measurements were performed within the specified temperature range of use of the refractive index liquids, *i.e.* from 18°C to 33°C. Fiber A was first considered. The thermal sensitivity of the FBG in the empty fiber, noted $S_{FBG} = \Delta\lambda_B/\Delta T|_{FBG}$ is 9.6pm/°C. That of the FBG in the fiber filled with an index liquid is simply noted $S=\Delta\lambda_B/\Delta T|_{device}$. It obviously depends on $S_{FBG}$, but also on the sensitivity of $\lambda_B$ to the thermal variations of the effective index of the mode noted $S_{neff}=\Delta\lambda_B/\Delta T|_{neff}$, that is negative. As shown on Fig. 4, the sensitivity S decreases as $n_{ref}$ increases. This is due to the fact that the extent of the guided mode is more significant when the refractive index value of the inserted medium is higher. This means that the influence of the changes of the liquid index with respect to temperature becomes stronger when the liquid inserted into the holes presents a higher refractive index value. Using Liquid 3 with $n_{ref} = 1.42780$ in Fiber A, S is divided by a factor ~ 3, the shift of $\lambda_B$ being limited to only 45 pm over

15°C instead of 140 pm for the empty MOF.

The index liquid with $n_{ref}$ = 1.4398 (Liquid 4) allows to cancel the sensitivity S of the FBG in Fiber A between 24.5°C and 26.5°C, the guided mode being less confined than in previous experiences (Fig. 5). At higher temperature, the refractive index of the medium being decreased, the spread of the guided mode into the holes is reduced. Under these conditions, the effect of the effective index of the mode is not sufficient to compensate for the influence of the temperature on silica. That is the reason why S remains positive over 27°C. On the opposite, at temperatures lower than 24°C, the light largely extends into the holes and the effects of the variations of the refractive index of the medium become predominant. The slope S then becomes negative

The same experiment was performed with Fiber B (Fig. 6). With Liquid 5, the sensitivity is cancelled around 22°C. The reference index of this liquid is slightly higher than that used to cancel the sensitivity in Fiber A because the field is more confined in Fiber B than in Fiber A, due to the larger holes of the cladding. The maximum absolute slope is limited to about 5 pm/°C to 6 pm/°C near 18°C and 30°C. The extrapolation of the curve depicted on Fig. 6 towards low temperatures shows that the variations of $\lambda_B$ may be kept within -25pm to +25pm over about 20°C (from 12°C to 33°C) whereas these variations are approximately ± 100 pm in the empty MOF over the same temperature range. So, the total range is reduced from 200 pm to 50 pm, *i.e.* a four-fold reduction is obtained.

### IV.  NUMERICAL VALIDATION OF THE METHOD

In order to evaluate the potential capability of the proposed technique, we have performed some numerical calculations the shift of $\lambda_B$ in different MOFs filled with various liquids. The refractive index of the pure silica is 1.444 at 1550nm and 25°C with a thermal dependence coefficient of $8 \cdot 10^{-6}$ °C$^{-1}$ [17] and a dilatation coefficient equal to $5.5 \cdot 10^{-7}$ [18,19]. At each temperature, the Bragg wavelength given by $\lambda_B(T) = 2 \cdot n_{eff}(T) \cdot \Lambda(T)$ must be calculated. The effective index of the fundamental mode is computed by means of a software based on the finite element method (FEM). It consists in first carefully drawing the cross section of the MOF. Then, each region of the structure, i.e. the holes, the doped core and the pure silica region, is finely meshed into elementary subspaces and the Maxwell equations are solved at each node of the mesh [20]. The period of the grating at 25°C is assumed to be such that $\lambda_B(25°C)=1550nm$. The shift of $\lambda_B$ in Fiber A filled with air, and with liquids 1 to 3 was first calculated. As shown in Fig. 7, the results are in very good agreement with those obtained experimentally (see Fig. 4). Then, Fiber A was filled with liquid 4 (Fig.8). We numerically found that the sensitivity of the FBG was cancelled with this liquid but the cancellation was obtained around 30°C instead of 25°C in the experiment. This discrepancy can be explained by the fact that the effective index of the mode is more and more sensitive to the geometrical parameters of the fiber, as the index of the medium is increased. Thus, in this preliminary study, we have modeled a ideal MOF having the mean geometrical parameters of Fiber A instead of its actual cross section. The cancellation of the sensitivity is numerically obtained at 25°C, with liquid 4, in a ideal MOF with d=2.9µm and D=6.5µm (Fig. 8). We can also note that the increase of the slope of the theoretical curve on both sides of the cancellation temperature is slightly lower than that of the experimental curve.

At last, we have searched for a MOF in which the Bragg wavelength is stabilized over a broadened range of temperatures. In Fig. 9, we show the example of a MOF with d=3µm and D=4.4µm, having the same doped core as Fiber A ( diameter = 5µm ; refractive index 8. $10^{-3}$ higher than that of pure silica). The fiber is filled with a liquid having a refractive index at 25°C $n_{ref}$ =1.39. The considered range of temperature is broadened from 0°C to 50°C. The thermal sensitivity of the liquid is assumed to remain equal to $-4$. $10^{-4}$ $°C^{-1}$ over the whole range of temperature. The variations of $\lambda_B$ remain within -12pm to +12pm over the whole range of temperature. The maximum slope is $-2$pm/°C near 0°C and +1.7pm/°C near 50°C. These very first calculations show that the Bragg wavelength of a FBG photowritten in a MOF can be stabilized over a significant range of temperature, provided that an adequate medium with a negative thermal sensitivity over this range of temperature can be found and inserted into the holes.

## V. CONCLUSION

In this paper, we have experimentally shown that it is possible to significantly reduce the thermal variations of the resonance wavelength $\lambda_B$ of FBGs photowritten in air/silica MOFs by inserting into the holes an index liquid with suitable reference value and appropriate thermal dependence. Based on this method, the range of variations of $\lambda_B$ has been reduced by a factor of 4 over 20°C in a 6-hole MOF. These first promising results have been confirmed by a  preliminary numerical approach based on the accurate calculation of the effective index of the guided mode using a finite element method (FEM). The numerical results are in very good accordance with the measurements. Furthermore, our computations show that the thermal variations of $\lambda_B$ can be significantly reduced over a broadened range of temperature in MOFs. For example, these variations remain within $-12$ pm to $+12$pm over 50°C for a FBG photowritten in a realistic MOF (d=3µm and D=4.4µm) filled with a medium having its refractive index equal to 1.39 at 25°C and its thermal sensitivity equal to $-4.10^{-4}$ $°C^{-1}$. By coupling such calculations with genetic algorithms [21], optimized geometrical parameters of MOFs and adequate media to be inserted into the holes could be identified,  in order to stabilize $\lambda_B$ over a even more broadened temperature domain. This extensive study is currently under investigation and will be published in a next paper.

Such passively stabilized FBGs present a great interest for accurate sensing of many parameters, e.g. strain and pressure,

and also for applications requiring stable narrow bandwidth fiber Bragg mirrors, such as wavelength division multiplexing (WDM) devices for telecommunication systems and fiber lasers.

### ACKNOWLEDGMENT

The authors are very grateful to J.L. Auguste, J.M. Blondy, P.O. Martin and P. Roy who manufactured the fibers. They also gratefully thank Nicolas Mothe for his help in implementing the software for the numerical calculations.

## REFERENCES

[1]  G. Meltz, W.W. Morey, and W.H. Glenn, "Formation of Bragg gratings in optical fibers by a transverse holographic method", *Opt. Lett*., 4, 823-825 (1989)

[2]  A. Othonos, and K. Kali, *Fiber Bragg Gratings, Fundamentals and Applications in Telecommunications and Sensors*, London, Artech House, 1999

[3]  J. Matsuoka, N. Kitamura, S. Fujinaga, T. Kitaoka, and H. Yamahita, "Temperature-dependence of refractive-index od $SiO_2$ glass", *J. Non-Cryst. Solids*, 135, 86-89 (1991)

[4]  M.J.N. Lima, R.N. Nogueira, J.C.C. Silva, A.L.J. Teixeira, P.S.B. André, J.R.F. da Rocha, H.J. Kalinowski and J.L. Pinto,"Comparison of the temperature dependence of different types of Bragg gratngs", *Mic. Opt. Tech. Letters*, 45 (4), 305-307 (2005).

[5]  B.J. Eggleton, P.S. Westbrook, R.S. Windeler, S. Spalter and T.A. Strasser, "Grating resonances in air-silica microstructure fibers," *Opt. Lett.* 24(21), 1460-1462 (1999).

[6]  J. C. Knight, T. A. Birks, P. S. J. Russell, and D. M. Atkin, "All-silica single-mode optical fiber with photonic crystal cladding," *Opt. Lett.* 21(19), 1547-1549 (1996)

[7]  J. Broeng, D. Mogilevstev, S.E. Barkou, and A. Bjarklev, "Photonic Crystal Fibers: A new class of optical waveguides," *Opt. Fib. Technol*. **5**(3), 305-330, (1999).

[8]  T.M. Monroe, W. Belardi, K. Furusawa, J.C. Baggett, N.G.R. Broderick, and D.J. Richardson, "Sensing with microstructured optical fibres," *Meas. Sci. and Tech.* 12(7), 854-858, (2001).

[9]  M.C. Phan Huy, G. Laffont, Y. Frignac, V. Dewynter-Marty, P. Ferdinand, P. Roy, J.-M. Blondy, D. Pagnoux, W. Blanc and B. Dussardier, "Fibre Bragg grating photowriting in microstructured optical fibres for refractive index measurement," *Meas. Sci. and Tech.* 17(5), 992-997, (2006).

[10] M.C. Phan Huy, G. Laffont, V. Dewynter-Marty, P. Ferdinand, L. Labonté, D. Pagnoux, P. Roy, W. Blanc and B. Dussardier, "Tilted Fiber Bragg Grating photowritten in microstructured optical fiber for improved refractive index measurement," *Opt. Express* 14(22), 10359-10370, (2007).

[11] C. Martelli, J. Canning, N. Groothoff and K. Lyytikainen, "Strain and temperature characterization of photonic crystal fiber gratings," *Opt. Lett.* 30(14), 1785-1787, (2005).

[12] G.W. Yoffe, P.A. Krug, F. Ouellette and D.A. Thorncraft, "Passive temperature-compensating package for optical fiber gratings," *Appl. Opt.* 34 (30), 6859-6861 (1995).

[13] Y. Huang, J. Lie, G. Kai, S. Yuan, and X. Dong, "Temperature compensation package for fiber Bragg gratings", *Mic. Opt. Tech. Lett.*, 39, 70-72 (2003)

[14] D.L. Weidman, G.H. Beall, K.C. Chyung, G.L. Francis, R.A. Modavis, and R.M. Morena, "A novel negative expansion substrate material for athermalizing fiber Bragg gratings" *$22^{nd}$ European Conference on Optical Communication (ECOC'96, Oslo)*, MoB3.5, 1.61-1.64, (1996)

[15] T. Iwashima, A. Inoue, M. Shigematsu, N. Nishimura and Y. Hattori, "Temperature compensation technique for fibre gratings using crystalline polymer tubes," *Electron. Lett.* 33(5), 417-419 (1997).

[16] J.M. Lourtioz, and H. Rigneault, *Nanophotonics*. Ed. international scientific and  technical encyclopedia (ISTE), 2006, ch..3 "Photonic crystal fibers"

[17] J. H. Wray and J.T. Neu, "refractive index of several glasses as a function of wavelength and temperature", *J. Opt. Soc. Am.*, 59(6), 774-776 (1969)

[18] internet site : http://www.accuratus.com/fused.html

[19] S.H. Kim, C.K. Hwangbo, "Temperature dependence of transmission center wavelength of narrow bandpass filters prepared by plasma ion-assisted deposition", *J. Kor. Phys. Soc*, 45(1), 93-98 (2004)

[20] F. Bréchet, J. Marcou, D. Pagnoux and P. Roy, "Complete analysis of the propagation characteristics into photonic crystal fibers by the finite element method", ***Opt. Fib. Tech.,*** 6(2), 181-191 (2000)

[21] E. Kerrinckx, L. Bigot, M. Douay, and Y. Quiquempois, "Photonic crystal fiber design by means of a genetic algorithm," *Opt. Express,* 12 (9), 1990-1995 (2004)

TABLE I
REFRACTIVE INDICES OF THE MEDIA INSERTED INTO THE HOLES OF THE FIBERS

| Medium | Index $n_{ref}$ at 25°C and 1550nm | Thermal sensitivity $dn_l / dT$ ($10^{-4}$ °C$^{-1}$) |
|---|---|---|
| Air | 1 | 0 |
| Liquid 1 | 1.38970 | -4.12 |
| Liquid 2 | 1.41576 | -4.01 |
| Liquid 3 | 1.42780 | -3.96 |
| Liquid 4 | 1.43984 | -3.97 |
| Liquid 5 | 1.44580 | -3.96 |

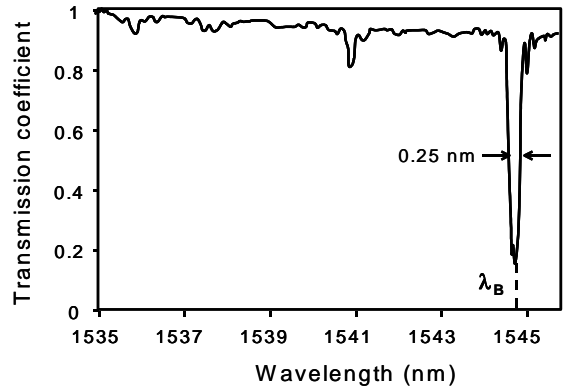

Fig. 2. Typical transmission spectrum of a FBG written in Fiber A; Full Width at Half Maximum of the peak = 0.25nm

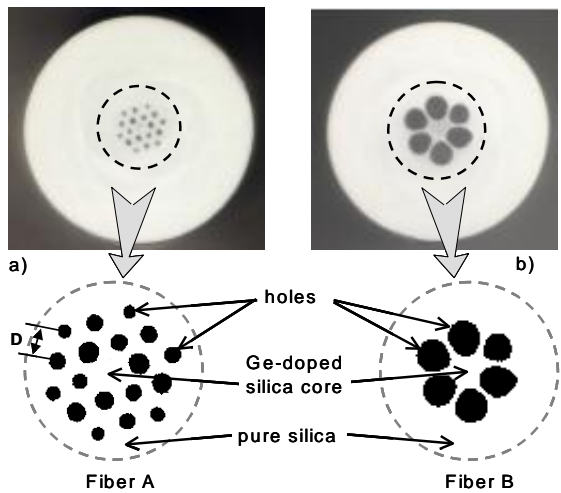

Fig. 1. Cross section of the 18-hole-MOF ="Fiber A" (a), and of the 6-hole-MOF ="Fiber B" (b). D is the distance between the centers of neighboring holes denoted the pitch

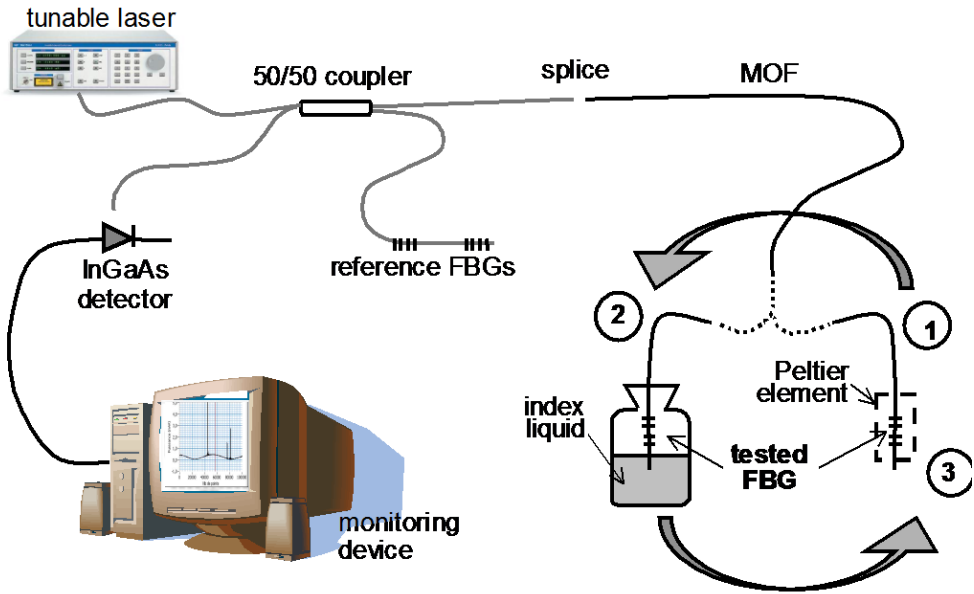

Fig. 3. Experimental setup for the measurement of the $\lambda_B$ shift of an FBG photowritten in a MOF, *vs* temperature. Configuration ①: measurement of the empty MOF. Configuration ②: Filling of the MOF with an index liquid. Configuration ③: measurement of the MOF filled with the index liquid

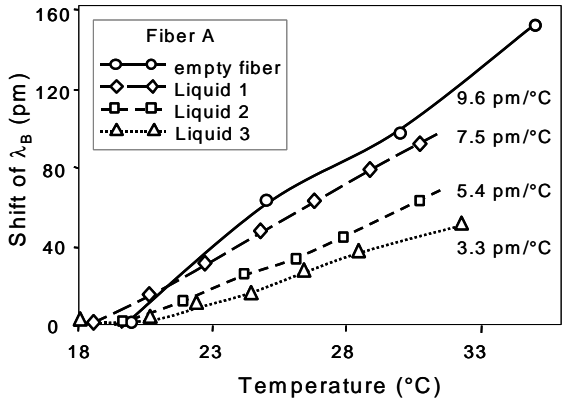

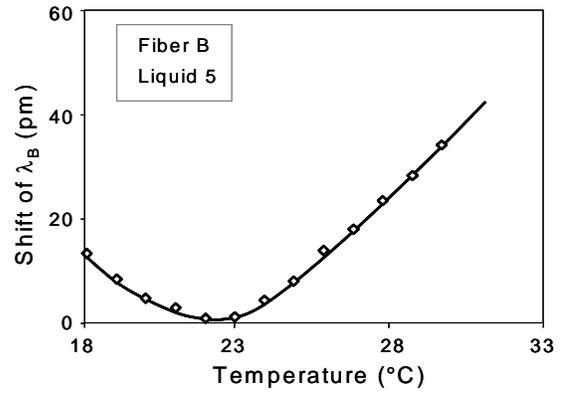

Fig. 4. Shift vs temperature for an FBG photowritten in Fiber A, measured for different index liquids,. The mean slope is indicated on the right.

Fig. 6: Resonance wavelength shift of a FBG photowritten in Fiber B, with a proper matching refractive index liquid ($n_{eff}$ = 1.4458) inserted into its holes.

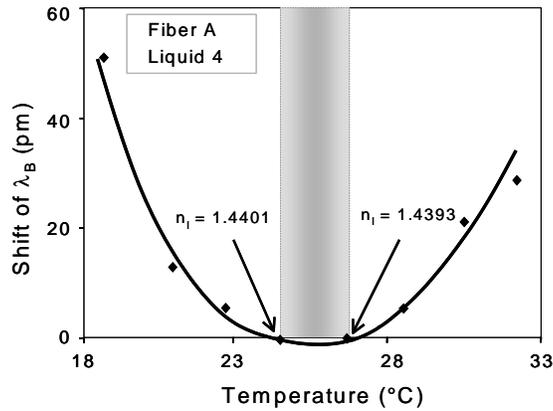

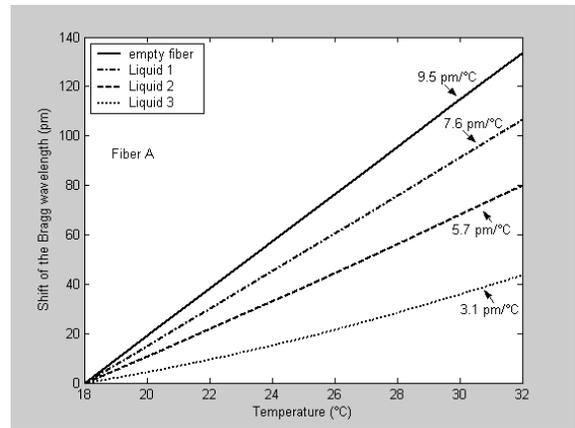

Fig. 5: Resonance wavelength shift of a FBG photowritten in Fiber A, with a proper matching refractive index liquid ($n_{ref}$ = 1.43984) inserted into its holes. $\Delta\lambda_B/\Delta T|_{neff}$ prevails under 24°C whereas $\Delta\lambda_B/\Delta T|_{FBG}$ prevails above 26°C. Between 24°C and 26°C (shaded region) the two effects are compensated.

Fig. 7: Shift of $\lambda_B$ vs temperature calculated for an FBG photowritten in Fiber A, for different index liquids. The means slope is indicated on the right

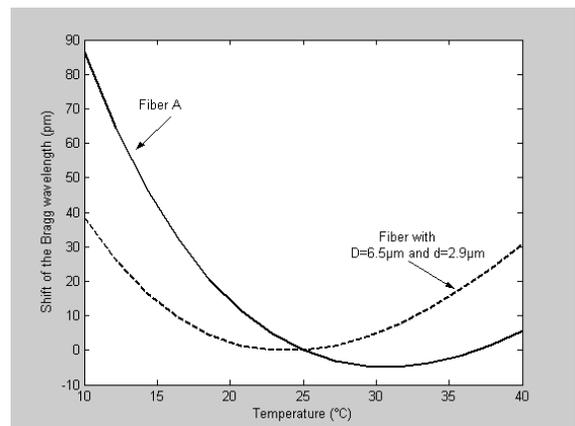

Fig. 8: Shift of $\lambda_B$ vs temperature calculated for an FBG photowritten in Fiber A and in a fiber with slightly smaller holes, filled with liquid 4

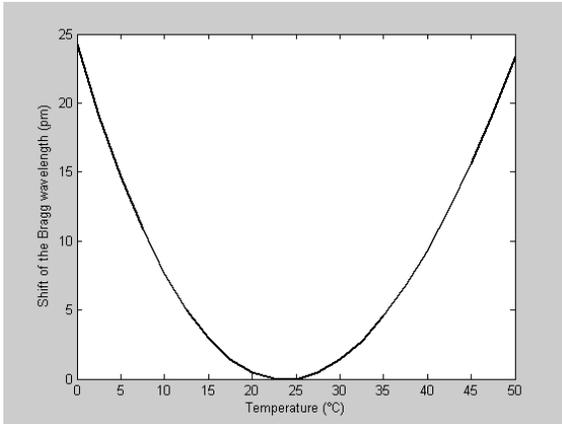

Fig. 9: Shift of $\lambda_B$ vs temperature calculated for an FBG photowritten in a MOF with D=4.4µm and d=3µm, filled with a medium having a refractive index at 25°C =1.39 and a thermal sensitivity =$4.10^{-4\circ}C^{-1}$